\documentclass[12pt]{iopart}
\usepackage{graphicx}
\usepackage{amssymb}

\begin{document}

\title[Photon condensation in circuit QED by engineered dissipation]{Photon condensation in circuit QED by engineered dissipation}
\author{D Marcos$^1$, A Tomadin$^1$, S Diehl$^{1,2}$ and P Rabl$^1$
}
\address{$^1$Institute for Quantum Optics and Quantum Information of the Austrian Academy of Sciences, A-6020 Innsbruck, Austria}
\address{$^2$Institute for Theoretical Physics, University of Innsbruck, A-6020 Innsbruck, Austria}
\ead{david.marcos@uibk.ac.at}

\date{\today}

\begin{abstract}
  We study photon condensation phenomena in a driven and
  dissipative array of superconducting microwave resonators.  Specifically, we
  show that by using an appropriately designed coupling of microwave photons to
  superconducting qubits, an effective dissipative mechanism can be engineered,
  which scatters photons towards low-momentum states while conserving their
  number. This mimics a tunable coupling of bosons to a low
  temperature bath, and leads to the formation of a stationary photon condensate
  in the presence of losses and under continuous-driving conditions. Here we
  propose a realistic experimental setup to observe this effect in two or
  multiple coupled cavities, and study the characteristics of such an
  out-of-equilibrium condensate, which arise from the
  competition between pumping and dissipation processes.
\end{abstract}

\maketitle

\section{Introduction}

Massive particles and spin systems have been at the center of
  investigations in quantum many-body physics since the inception of the field. 
  Photonic systems, however, have been the object of
  interest in this field only much more recently.
  This is, on the one hand, due to the fact that photons
are intrinsically non-interacting, and on the other hand, due to the difficulty
to confine and experiment with photons before they decay.  One important
exception in this context are ideas to produce a Bose-Einstein condensate (BEC)
of weakly interacting polaritons \cite{DHY2010,DWSBY2002,KRKBJKMSASSLDD2006,BHSPW2007, ASLBVMLBKSTV2009}, 
whose large photonic component is essential to
achieve a low effective mass and therefore transition temperatures which are
accessible in solid-state experiments.  While current cold-atom experiments
offer a well controlled setting to study bosonic quantum gases \cite{BDZ2008}, the
investigation of photonic condensates is still an active area of research,
stimulated by new experimental directions in the field \cite{KVM2010,KSVM2010,KSDVW2011} and controversial discussions on the
exact nature of such condensates \cite{Snoke2003,SL2002,SLS2003}.

More recently, the connection between many-body physics and photons has been
addressed again from the perspective of photonic quantum
simulators~\cite{HBP2008,TF2010,HBP2006,GTCH2006,ASB2007,RF2007,NUTY2008,FOU2008,KH2009, SGHBT2010}.  Here, the general goal is to implement
\emph{strongly} interacting many-body systems in a controllable way, to simulate
the properties of non-trivial condensed-matter
models.  This can be achieved, for example,
using ideas from cavity quantum electrodynamics (QED) \cite{RBH2001}, 
where effective photon-photon interactions can be obtained through the coupling to an
intermediary system, such as atoms \cite{I1997,BBMBNK2005} or solid-state systems \cite{HBWGAGFHI2007,FFESPV2008,RTM2009}.  Based on this
principle, various schemes for implementing bosonic Hubbard
models for photons on a lattice~\cite{HBP2008,TF2010}, photonic quantum Hall
systems~\cite{CAB2008,Koch2010,Hafezi2011,UC2011}, and strongly interacting photons in a 1D
continuum~\cite{CGMVLD2008,CGTDCI2009}, have been theoretically investigated.  Although the
experimental implementation of these ideas is still challenging, the development
of scalable cavity-QED systems in on-chip devices \cite{LTHPH2011} is rapidly progressing, and
analogous (`circuit QED') systems in the microwave regime \cite{WSBFHMKGS2004, CBSNHM2004,BHWGS2004,SG2008,CW2008,MWBLLNCSWWYYZMC2011} 
already approach the stage where photonic lattice models can be realized.

A central and still open question in the field of photonic quantum simulators
is how to prepare and probe quantum many-body states of light.  Because of
unavoidable losses and the absence of a chemical potential (arising from an
equilibrium particle reservoir), photonic many-body systems must necessarily be
studied under driven and non-equilibrium conditions. Therefore, familiar
concepts from condensed-matter physics, such as ground-state or equilibrium phase
diagrams, are {\it a priori} not accessible in photonic many-body systems.  In
previous works it has been suggested to study, for example, the transient
dynamics of an initially pumped system~\cite{HBP2006,ASB2007,TGFGCTI2010}, or to use excitations
spectroscopy and photon-correlation measurements of a weakly driven system to
infer certain properties of the interacting photonic
system~\cite{CGTDCI2009,PEH2009,KAL2010,LH2010,KAWLC2011}.

In this work we consider a different scenario and study the dissipative dynamics
of a \emph{strongly} driven cavity array in the presence of engineered
dissipation \cite{Poyatos96,Diehl2008,Verstraete09,WMLZ2010,Kastoryano11,Muschik11,Barreiro11,Krauter11}.  More precisely, we will show how in a circuit QED setting, the
coupling of microwave photons to superconducting qubits can be used to scatter
photons from high to low momentum states, eventually accumulating photons in a
zero-momentum condensate. Previously, a similar scheme has been explored as a
dissipative way to prepare a Bose Einstein condensate (BEC) of
atoms~\cite{Diehl2008}. In the case of photonics systems, it is essential that
this mechanism implements a \emph{number-conserving} coupling of photons to an
effective low-temperature bath, and therefore provides a new approach for the
preparation of quasi-equilibrium many-body states in open and driven photonic
systems.  Here we propose and analyze a proof-of-principle experiment to study
condensation of microwave photons in a dissipative cavity array, and describe
the properties of such a non-equilibrium BEC, which arise from the interplay
between driving, decay, and thermalization processes. 
Such experiment would realize a controlled setting for the simulation of 
non-equilibrium condensation phenomena, and more generally, would offer a new route
towards preparing stationary states of strongly interacting photons. 
Moreover, our work shows that the flexibility provided by circuit QED, which
so far has mainly been employed to engineer strong coherent interactions, 
can equally well be used for the design of various non-trivial dissipative couplings, and
be applied to simulate driven quantum systems in unconventional environments.

\section{Coupled cavity arrays}\label{cavityarrayintro}

\begin{figure}
\begin{center}
\includegraphics[width=0.8\linewidth]{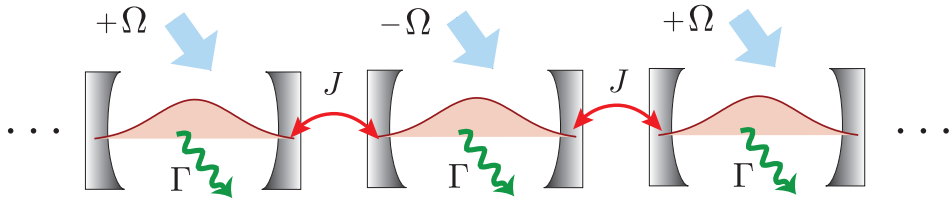}
\caption{\label{cavity_array}
Schematics of a coupled array of photonic cavities (represented in grey).
The photons in each cavity are tunnel-coupled to neighboring sites with amplitude $J$ and decay with rate $\Gamma$.
The cavities are driven with an external coherent field of strength $|\Omega|$. This external driving compensates losses 
and ensures a finite stationary photon population.}
\end{center}
\end{figure}

Fig.~\ref{cavity_array} illustrates the basic setup for an array of $L$ coupled
cavities, where each cavity is represented by a single photonic mode of
frequency $\omega_{\rm c}$ and bosonic annihilation operator $\hat c_\ell$.  Photons can
tunnel between neighboring cavities with hopping amplitude $J$ and, in addition,
local interactions with two- or many-level systems can be used to induce
Kerr-type nonlinearities with an effective photon-photon interaction strength
$U$.  A generic model for this system is then given by the Bose-Hubbard
Hamiltonian (see e.g. Ref.~\cite{HBP2008})
\begin{equation} \label{BHmodel}
\hat{\cal H}_{\rm c}= \sum_{\ell} \omega_{\rm c} \hat{c}_{\ell}^{\dag} \hat{c}_{\ell}  - J \sum_{\ell}  (\hat{c}_{\ell} \hat{c}_{\ell+1}^\dag +  \hat{c}_{\ell+1} \hat{c}_{\ell}^{\dag})  + \frac{U}{2}\sum_{\ell}  \hat{c}^{\dag}_{\ell} \hat{c}_{\ell}^\dag \hat{c}_{\ell} \hat{c}_{\ell}.
\end{equation}
Various generalizations of this model have been discussed in the literature and
several implementations have been proposed using optical or microwave
cavities. The \emph{photonic} Hubbard model given in Eq.~(\ref{BHmodel})
describes a gas of interacting bosons on a lattice, where the hopping $J$
competes with on-site interaction $U$. However, in contrast to the cold-atom
physics scenario, the bosons here can decay, and under experimentally-relevant
conditions $\omega_{\rm c} \gg J,U,k_{\rm B}T$ (where $T$ is the
temperature and $k_{\rm B}$ the Boltzmann constant), the equilibrium state of
this model is simply the vacuum state. Therefore, in photonic many-body systems
we are mainly interested in the out-of-equilibrium dynamics of $\hat{\cal
  H}_{\rm c}$ in the presence of losses and external driving fields. In
particular, in this work we model the resulting dissipative dynamics for the
system density operator $\rho$ by a master equation (ME) of the form,
\begin{equation}\label{effectivecavity} 
\dot \rho = - i[\hat{\cal H}_{\rm c} + \hat{\cal H}_\Omega(t),\rho] + \Gamma \sum_\ell \mathcal{D}[\hat c_\ell]\rho +{\cal L}_{\kappa} \rho,
\end{equation}
where $\mathcal{D}[\hat{c}]\rho \equiv 2 \hat{c} \rho \hat{c}^\dag -
\hat{c}^\dag \hat{c} \rho - \rho \hat{c}^\dag \hat{c}$. In
Eq.~(\ref{effectivecavity}) the Hamiltonian $\hat{\cal H}_{\Omega}(t)=
\sum_{\ell} \Omega_{\ell} ( e^{-i \omega_{\rm d} t} \hat{c}_{\ell}^{\dag} + e^{i
  \omega_{\rm d} t} \hat{c}_{\ell})$ describes an external driving field of
frequency $\omega_{\rm d}$ which is used to excite the system, and the second
term accounts for photon losses in each cavity with a field decay rate $\Gamma$.
While a finite driving field is required to counteract the losses, it will in
general also compete with $\hat{\mathcal{H}}_{\rm c}$ and, for strong driving
fields, even dominate the system dynamics. Therefore, in previous works it has
been suggested to study either the transient dynamics of an initially prepared
photonic state \cite{HBP2006,ASB2007,TGFGCTI2010} (where $\Omega_{\ell}=0$ for times $t>0$)~\cite{TGFGCTI2010} or
to use weak excitation spectroscopy \cite{CGTDCI2009,PEH2009,KAL2010,LH2010,KAWLC2011} ($\Omega_{\ell}\rightarrow 0$) to probe the
many-body spectrum of Hamiltonian $\hat{\cal H}_{\rm
  c}$.
 
In this work we are interested in the opposite regime of a \emph{strongly} and
\emph{continuously} driven cavity array. We study the dynamics of this system in
the presence of an additional artificial thermalization mechanism, denoted by
${\cal L}_{\kappa}$ in Eq.~(\ref{effectivecavity}). More precisely, we will
show below how a non-local coupling of photons to superconducting qubits can be
engineered in an array of microwave cavities to implement a dissipative photon
scattering process of the form
\begin{equation} \label{Lth}
{\cal L}_{\kappa} =  \sum_{\ell} \frac{\kappa}{4} \mathcal{D}[(\hat{c}_{\ell}^\dag + \hat{c}^{\dag}_{\ell+1})(\hat{c}_{\ell} - \hat{c}_{\ell+1})]  + \frac{\kappa'}{4} \mathcal{D}[(\hat{c}_{\ell}^\dag - \hat{c}_{\ell+1}^{\dag})(\hat{c}_{\ell} +\hat{c}_{\ell+1})].
\end{equation}
The interpretation of this term can be seen best in the case of just two
cavities. Then, the first term in Eq.~(\ref{Lth}) describes the scattering of
photons from the asymmetric (energetically higher) mode $\hat{c}_a \equiv
(\hat{c}_1 - \hat{c}_2)/\sqrt{2}$ into the symmetric (energetically lower) mode
$\hat{c}_s \equiv (\hat{c}_1 + \hat{c}_2)/\sqrt{2}$, while preserving the total
photon number.  The second term in Eq.~(\ref{Lth}) describes the reverse
process. In the absence of losses, the action of ${\cal L}_{\kappa}$ would, for two cavities,
drive arbitrary initial photon states into a thermal equilibrium distribution, 
characterized by a detailed balance between symmetric and antisymmetric modes,
with the ratio $\kappa'/\kappa= \exp(-2 J/k_{\rm B}T_{\rm eff})$ defining
an effective temperature $T_{\rm eff}$. Similarly, for a whole cavity array, the
processes $\sim \kappa$ scatter photons to lower momenta and -- as
previously shown for an analogous system of cold atoms~\cite{Diehl2008} -- drive the system
towards a condensate of photons in the zero momentum mode, $|{\rm BEC} \rangle =
(\hat{c}_{q=0}^{\dag})^N |0\rangle / \sqrt{N!}$, where $q$ is the quasimomentum
and $N$ the total number of particles.  Also in this case the opposite processes
$\sim \kappa'$ can be roughly interpreted as a finite temperature effect, although, as it will be shown below, for
$L>2$ and $\kappa'\neq0$ the stationary state of ${\cal L}_{\kappa}$ does no
longer represent a true thermal equilibrium state.
 
The competition of ${\cal L}_{\kappa}$ with a Kerr-type nonlinearity in the
Hamiltonian has already been discussed in the atomic, number-conserving
setup~\cite{DTMFZ2010,Tomadin11}.  However, in the photonic case, under realistic
conditions, ${\cal L}_{\kappa}$ competes with external driving fields and
intrinsic photon losses as described by the full ME~(\ref{effectivecavity}).  To
study this competition more explicitly, we focus in the following on the limit
$U \rightarrow 0$, and consider the experimental scenario where neighboring
cavities are driven by coherent fields of alternating amplitude
$\Omega_\ell=(-1)^\ell \Omega$ (see Fig.~\ref{cavity_array}). In this case, the
accumulation of photons in the $q=0$ mode can be interpreted as a clear
signature for photon condensation induced by the dissipative photon-photon
interactions in ${\cal L}_{\kappa}$.

\section{Physical implementation}

\begin{figure}
\includegraphics[width=0.48\linewidth]{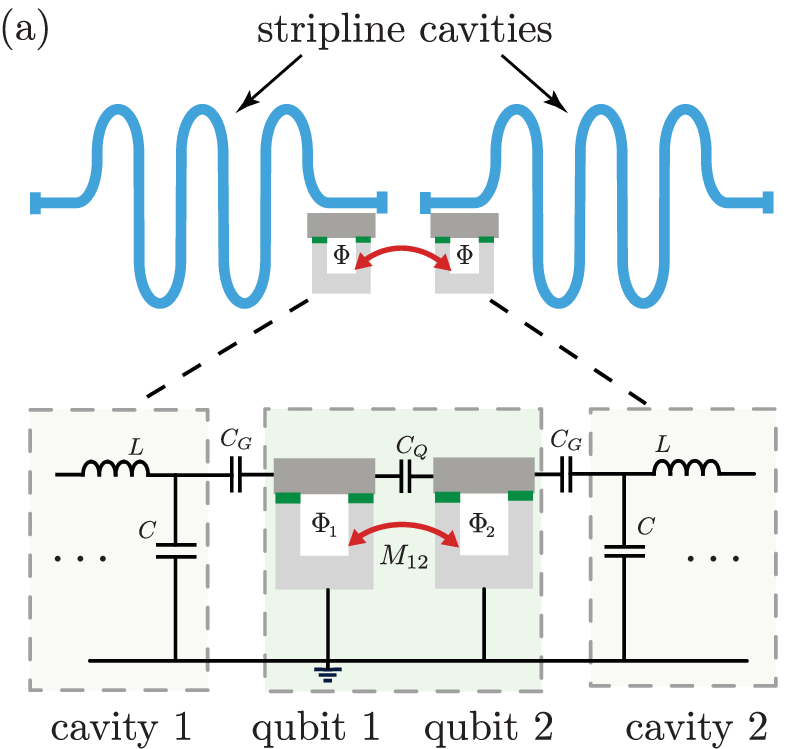}
\includegraphics[width=0.48\linewidth]{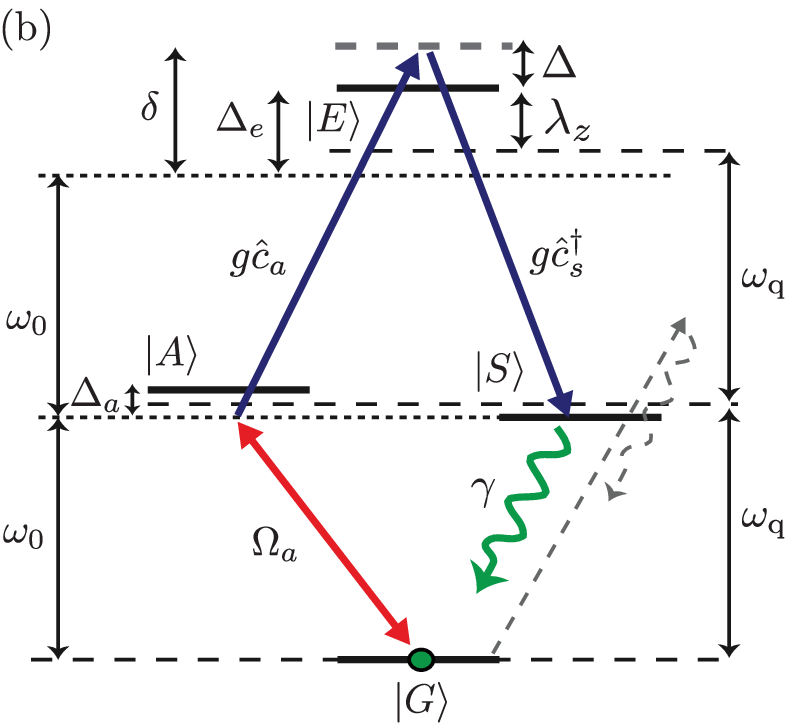}
\caption{\label{circuit} Dissipation engineering with circuit QED. 
a) Implementation of the Hamiltonian in Eq.~(\ref{initialHamiltonian}).
 Two stripline resonators are coupled through a
  system of two mutually interacting Cooper pair boxes.  b) Four-level diagram
  corresponding to the two coupled qubits, whose dynamics is described by
  Eq.~(\ref{rotatingframeH}). }
\end{figure}

In this section we show how the photon scattering mechanism described by
Eq.~(\ref{Lth}) can be implemented in an array of superconducting microwave
resonators.  The basic setting is illustrated in Fig.~\ref{circuit}a for two
cavities, each coupled to a nonlinear element, for example a superconducting
Cooper pair box (`charge qubit').  The qubits are placed next to each other to
obtain strong electrostatic or magnetic interactions.  As we discuss now, this
configuration allows us to engineer both nonlinear as well as non-local
dissipation processes for photons.

For simplicity we restrict the following analysis to a single block of only two
cavities, but the generalization to a whole array of linked
cavities is straightforward. The Hamiltonian for this system can be derived from
the corresponding equivalent circuit model schematically shown in
Fig.~\ref{circuit}a. By restricting each resonator to a single mode
$\hat{c}_{\ell}$, $\ell=1,2$, and by approximating each Cooper pair box by a
two-level system with ground state $|g\rangle$ and excited state $|e\rangle$, we
obtain
\begin{equation} \label{initialHamiltonian}
\hat{\cal H} = \sum_{\ell=1,2}  \omega_{\rm c} \hat{c}_{\ell}^\dag \hat{c}_{\ell} 
+ \sum_{\ell=1,2}  g (\hat{c}_{\ell}^\dag \hat\sigma_{-}^{(\ell)} + \hat{c}_{\ell} \hat\sigma_{+}^{(\ell)}) +  \hat{\cal H}_{\rm q} (t),
\end{equation}
where $\hat\sigma_{-}^{(\ell)} \equiv | g_{\ell} \rangle \langle e_{\ell} |$.  The
first part is the free cavity Hamiltonian, while the second
term describes the dipole interaction between photons and qubits with strength
$g$.  Finally, $\hat{\cal H}_{\rm q}(t)$ is the Hamiltonian of the two coupled
qubits, which we assume of the generic form \footnote{The explicit derivation using circuit theory is omitted here. For a comprehensive review on the topic c.f. \cite{BHWGS2004,D97,MSS2001}.}
\begin{eqnarray} \label{qubitH}
\hat{\cal H}_{\rm q}(t)&=& \sum_{\ell=1,2}[ \omega_{\rm q} |e_{\ell} \rangle \langle e_{\ell}| + \Omega_{\rm q}^{(\ell)} ( e^{-i\omega_0 t} \hat\sigma_+^{(\ell)} + e^{i\omega_0 t} \hat\sigma_{-}^{(\ell)}) ]\nonumber \\
& &+ \lambda_z \hat\sigma_{ee}^{(1)} \hat\sigma_{ee}^{(2)} + \lambda_x (\hat\sigma_{+}^{(1)} \hat\sigma_{-}^{(2)} + \hat\sigma_{-}^{(1)} \hat\sigma_{+}^{(2)}).
\end{eqnarray}
Here, $\Omega_{\rm q}^{(\ell)}$ are the Rabi frequencies of an applied microwave
field, which is used to drive the qubits at frequency $\omega_0$.  The coupling
constants $\lambda_z$ and $\lambda_x$ are the strengths of the diagonal and
off-diagonal qubit-qubit interactions, respectively.  The
Hamiltonian~(\ref{initialHamiltonian}) can be rewritten in terms of the symmetric $\hat c_s \equiv (\hat{c}_1 +
\hat{c}_2)/\sqrt{2}$ and antisymmetric $\hat c_a\equiv (\hat{c}_1 - \hat{c}_2)/\sqrt{2}$ cavity modes, and the
qubit states $|E\rangle\equiv |ee\rangle$, $|G\rangle\equiv |gg\rangle$,
$|S\rangle\equiv (|eg\rangle+ |ge\rangle)/\sqrt{2}$, and $|A\rangle\equiv
(|eg\rangle- |ge\rangle)/\sqrt{2}$.  In this new basis, choosing $\Omega_{\rm
  q}^{(1)} = -\Omega_{\rm q}^{(2)} \equiv \Omega_a/\sqrt{2}$, and changing into a
frame rotating with $\omega_0$, we obtain
\begin{eqnarray} \label{rotatingframeH}
\hat{\cal H} & = & -\delta \hat{c}_s^\dag \hat{c}_s - \delta \hat{c}_a^\dag \hat{c}_a  - \Delta_e  |E\rangle\langle E|  - \Delta_s  |S\rangle\langle S| - \Delta_a |A\rangle\langle A| \nonumber \\
&& + \Omega_a [ |A\rangle \langle G| -  |E\rangle \langle A| + {\rm H.c.}] \nonumber \\
& & + g [ \hat{c}_s^\dag (  |G\rangle \langle S| + |S\rangle \langle E| ) + \hat{c}_a^\dag ( |G\rangle \langle A| - |A\rangle \langle E| ) + {\rm H.c.}],
\end{eqnarray}
where $\delta= \omega_0-\omega_{\rm c}$, $\Delta_{s,a} = \omega_0 - \omega_{\rm
  q} \mp \lambda_x$, $\Delta_e = 2(\omega_0 - \omega_{\rm q})-\lambda_z$.  Apart
from the coherent dynamics described by Eq.~(\ref{rotatingframeH}), we include
dissipation in the form of intrinsic cavity loss with rate $2\Gamma$ and qubit
decay with rate $\gamma$. The full system dynamics is then described by the ME
\begin{equation}\label{CircuitQEDME}
\dot \rho = -i[\hat{\cal H},\rho] + \sum_{\eta = s,a} \left (   \Gamma  \mathcal{D}[\hat{c}_\eta] \rho + \frac{\gamma}{2} \mathcal{D}[|G\rangle\langle \eta|] \rho + \frac{\gamma}{2} \mathcal{D}[|\eta \rangle\langle E|] \rho \right ).
\end{equation}
and a summary of the energy levels and the most relevant transitions is shown in
Fig.~\ref{circuit}b.

Our goal in the following is to eliminate the qubit dynamics and derive an
effective ME for the cavity modes.  Specifically, we are interested in the
dissipative two-photon process, where the qubits change from $|A\rangle$ to
$|S\rangle$ by absorbing a photon from the antisymmetric cavity mode ($\hat{c}_a$)
and emitting a photon into the symmetric cavity mode ($\hat{c}_s^{\dagger}$) (see
Fig.~\ref{circuit}b).  Since the overall process is $\mathcal{O}(g^4)$ in the
photon-qubit coupling strength, a general derivation is quite involved, and to
make the following discussion more transparent we will now focus explicitly on
the hierarchy of energy scales drawn in Fig.~\ref{circuit}b.  In
particular, we assume that $\delta$, $\Delta_e$, and $\Delta \equiv \delta-\Delta_e$
are much larger than all the other frequency scales $g, \Omega_a, \Delta_{s,a},
\Gamma, \gamma$.  In this limit, none of the single-photon processes is resonant
and we can use a Schrieffer-Wolff transformation to derive an effective two-photon Hamiltonian.  
We perform the unitary transformation $\tilde{\cal H}
\equiv \hat{V} \hat{\cal H} \hat{V}^\dag$ with $\hat{V}=e^{\hat{S}}$, and make
the ansatz
\begin{equation}
\hat{S} = \hat{c}_s^\dag  \left [ \vphantom{\sum}  \alpha_{1,s} |G\rangle \langle S| + \alpha_{2,s} |S\rangle \langle E| \right ]  + \hat{c}_a^\dag \left [ \vphantom{\sum} \alpha_{1,a} |G\rangle \langle A| - \alpha_{2,a} |A\rangle \langle E| \right ] - {\rm H.c.}~
\end{equation}
We define $\hat{\cal H}_0$ and $\hat{\cal H}_g$ as the first and last lines of
the Hamiltonian (\ref{rotatingframeH}), respectively, and choose the
coefficients $\alpha_i$ such that $[\hat{S},\hat{\cal H}_0] = -\hat{\cal H}_g$.
This can be achieved by setting
\begin{equation}
\alpha_{1,\eta} = \frac{g}{\delta-\Delta_\eta},
\qquad \alpha_{2,\eta}= \frac{g}{\delta-(\Delta_e -\Delta_\eta)},
\qquad \eta = s, a.
\end{equation}
In view of the assumptions discussed above, we will use the approximate results
$\alpha_{1,\eta}\approx g/\delta$, $\alpha_{2,\eta}\approx g/\Delta$ and define
the parameter $\epsilon=\Delta/\delta <1$.  After this transformation the
Hamiltonian (\ref{rotatingframeH}) reads
\begin{equation}
\tilde{\cal H} =  \tilde{\cal H}_{\rm q} + \tilde{\cal H}_{\rm c} + \tilde{\cal H}_{\rm int} + \mathcal{O}(g^3, g \Omega_a).
\end{equation}
Here $\tilde{\cal H}_{\rm q}$ is the qubit Hamiltonian $\hat{\cal H}_{\rm q}$
written above, with small frequency shifts absorbed into a redefinition of the
detunings $\Delta_{e,s,a}$.  The modified cavity Hamiltonian is
\begin{equation}
\tilde{\cal H}_{\rm c} = - \bar \delta \hat{c}_s^{\dag} \hat{c}_s - \bar \delta \hat{c}_a^{\dag} \hat{c}_a - J (\hat{c}_s^{\dag} \hat{c}_s - \hat{c}_a^{\dag} \hat{c}_a) + g_{\rm eff} \left[( \hat{c}_s^\dag \hat{c}_a+ \epsilon \hat{c}_a^\dag \hat{c}_s )  \langle P_{sa} \rangle + {\rm H.c.} \right],
\end{equation}
where $g_{\rm eff}\equiv g^2/\Delta$, $\bar \delta \equiv \delta - g_{\rm eff}
(\langle P_{aa} \rangle + \langle P_{ss} \rangle)/2$, $J\equiv g_{\rm eff}
(\langle P_{aa} \rangle- \langle P_{ss} \rangle)/2$, $P_{sa} \equiv
|S\rangle\langle A|$,
\begin{eqnarray}
P_{ss} & = &   (|S\rangle\langle S| - |E\rangle\langle E|) + \epsilon  (|G\rangle\langle G| -|S\rangle\langle S|), \\
P_{aa} & = &  (|A\rangle\langle A| - |E\rangle\langle E|) + \epsilon  (|G\rangle\langle G| -|A\rangle\langle A|),
\end{eqnarray}
and the average $\langle \cdot \rangle$ is taken with respect to the stationary
qubit state in the limit $g\rightarrow 0$.  For the parameter regime
of interest, $\langle P_{sa} \rangle\approx 0$, and $\tilde{\cal H}_{\rm c}$
corresponds to the free cavity Hamiltonian with a qubit-mediated tunneling
amplitude $J$.  Finally, the effective coupling between photons and qubits can
be written as
\begin{equation}
\tilde{\cal H}_{\rm int} \simeq  g_{\rm eff} \left [ \hat{c}_s^\dag \hat{c}_s \bar P_{ss} + \hat{c}_a^\dag \hat{c}_a  \bar P_{aa} +(\hat{c}_s^\dag \hat{c}_a+ \epsilon \hat{c}_a^\dag \hat{c}_s)  \bar P_{sa} + (\hat{c}_s^\dag \hat{c}_a+ \epsilon \hat{c}_a^\dag \hat{c}_s)^\dag \bar P_{sa}^\dag \right ],~
\end{equation}
where $\bar P_{km}=P_{km} -\langle P_{km}\rangle$.  Note that in $\tilde{\cal H}_{\rm
  int}$ we have only kept resonant two-photon processes and already omitted
terms like $\sim g\Omega_a/\Delta \times \hat{c}_{s,a}$ and $g^2/\Delta \times
\hat{c}_{s,a}^2$.  Both of these terms oscillate with the detuning $\delta$ and
will only give small corrections to the results presented below.

The transformation $\hat{V}$ induces a weak mixing between photon and qubit
states, which apart from generating new effective interactions also modifies the
dissipative couplings.  Since we consider $\gamma \gg \Gamma $, we find that the
main result of this mixing is an enhancement of the cavity decay rates $ \Gamma
\rightarrow \Gamma_{a,s} \equiv \Gamma + \tilde\Gamma_{s,a}$.  After a transformation of the jump operators in the ME~(\ref{CircuitQEDME}), and averaging over the qubit states, we find that for the parameter
regime of interest, these rates are
\begin{eqnarray}
\tilde\Gamma_{s}&\approx& \frac{\gamma g^2}{2\Delta^2} \epsilon^2 \left[ \langle |G\rangle\langle G| \rangle + \langle | A \rangle \langle A | \rangle \right], \\
\tilde\Gamma_{a}&\approx& \frac{\gamma g^2}{2\Delta^2} \left[ \epsilon^2 \langle |G\rangle\langle G| \rangle+ (2+\epsilon^2) \langle |A\rangle\langle A| \rangle \right].
\end{eqnarray}

In summary, we find that the system dynamics in the new dressed state
representation is well described by the ME
\begin{equation}
\dot \rho = ( \mathcal{L}_{\rm q} +  \mathcal{L}_{\rm c}  + \mathcal{L}_{\rm int} ) \rho,
\end{equation}
where
\begin{eqnarray}
\mathcal{L}_{\rm q} \rho = -i[\tilde{\cal H}_{\rm q},\rho] + \sum_{\eta=s,a} \left ( \frac{\gamma}{2} \mathcal{D}[|G\rangle\langle \eta|] \rho + \frac{\gamma}{2} \mathcal{D}[|\eta \rangle\langle E|] \rho \right ),\\
\mathcal{L}_{\rm c} \rho = -i [\tilde{\cal H}_{\rm c},\rho] + \sum_{\eta=s,a} \Gamma_{\eta} \mathcal{D}[\hat{c}_{\eta}] \rho,\\
\mathcal{L}_{\rm int}\rho = -i[\tilde{\cal H}_{\rm int},\rho].
\end{eqnarray}
In the limit $g_{\rm eff}\rightarrow 0 $, qubits and cavities are
decoupled, and the system relaxes on a timescale $\gamma^{-1}$ into the state
$\rho(t) \simeq \rho_{\rm c} (t) \otimes \rho_{\rm q}^0$, where
$\mathcal{L}_{\rm q} \rho_{\rm q}^0 = 0$.  For finite $g_{\rm eff} <\gamma$, we
can use a perturbative projection operator technique to eliminate the qubit degrees of
freedom and derive an effective ME for the reduced cavity density operator \cite{BP2002},
\begin{equation}
\dot \rho_{\rm c} = {\cal L}_{\rm c} \rho - \int_0^\infty d\tau \, {\rm Tr}_{\rm q}\{ [\tilde{\cal H}_{\rm int}, e^{\tau \mathcal{L}_{\rm q}} \left( [\tilde{\cal H}_{\rm int}, \rho_{\rm c}(t) \otimes \rho_{\rm q}^0] \right) ] \}.
\end{equation}
Evaluating this expression we obtain
\begin{eqnarray}\label{eq:EffCavityME}
\dot \rho_{\rm c} & \approx & -i[\tilde{\cal H}_{\rm c} + \hat{\cal H}_{\rm eff},\rho_{\rm c} ] + \sum_{\eta=s,a} \Gamma_{\eta} \mathcal{D}[\hat{c}_{\eta}] \rho_{\rm c} +   \kappa \mathcal{D}[(\hat{c}_s^\dag \hat{c}_a+ \epsilon \hat{c}_a^\dag \hat{c}_s)] \rho_{\rm c}  \nonumber \\
& &+ \sum_{\eta,\eta'=s,a}\Gamma^\Phi_{\eta\eta'} \left(\hat  n_{\eta} \rho_{\rm c} \hat n_{\eta'} + \hat n_{\eta'} \rho_{\rm c} \hat n_{\eta}
 - \hat n_{\eta} \hat n_{\eta'} \rho_{\rm c} - \rho_{\rm c} \hat n_{\eta} \hat n_{\eta'} \right ),
\end{eqnarray}
where $\hat n_\eta=\hat c_\eta^\dag \hat c_\eta$ and $\hat{\cal H}_{\rm eff}$ is
an effective photon-photon interaction
 \begin{equation}
\hat{\cal H}_{\rm eff}=  U_{\rm eff} (\hat{c}_s \hat{c}_a^\dag +\epsilon \hat{c}_a \hat{c}^\dag_s)(\hat{c}_s^\dag \hat{c}_a +\epsilon \hat{c}_a^\dag \hat{c}_s) +  \sum_{\eta,\eta'=s,a} U_{\eta\eta'} \hat n_{\eta} \hat n_{\eta'}.
\end{equation}
In Eq.~(\ref{eq:EffCavityME}) we have defined $\kappa \equiv g_{\rm eff}^2 {\rm
  Re} \{S_{as,sa}(0)\}$, $\Gamma^\Phi_{\eta\eta'} \equiv g_{\rm eff}^2 {\rm Re}
\{S_{\eta\eta,\eta'\eta'}(0)\}$, and the interactions $U_{\rm eff} \equiv
g_{\rm eff}^2 {\rm Im} \{S_{as,sa}(0)\}$ and $U^\Phi_{\eta\eta'} \equiv g_{\rm
  eff}^2 {\rm Im} \{S_{\eta\eta,\eta'\eta'}(0) \} $, given in terms of the qubit correlation spectra
\begin{equation} \label{spectrumEq}
S_{\eta\eta',\varsigma\varsigma'}(\omega) \equiv  \int_0^\infty d\tau \, e^{-i\omega \tau} \langle \bar P_{\eta\eta'}(\tau) \bar P_{\varsigma\varsigma'}(0)\rangle,
\end{equation}
which can be calculated using the quantum regression theorem \cite{GardinerZollerBook}.

\subsection{Discussion} 

By ignoring coherent photon-photon interactions for the moment, we see that the
first line of the effective cavity dynamics given in Eq.~(\ref{eq:EffCavityME})
represents the dissipatively coupled cavities as introduced in Eq.~(\ref{effectivecavity}). 
In particular, we can write
\begin{equation}\label{expandLth}
\kappa \mathcal{D}[(\hat{c}_s^\dag \hat{c}_a+ \epsilon \hat{c}_a^\dag \hat{c}_s)]  = \kappa  \mathcal{D}[\hat{c}_s^\dag \hat{c}_a] + \kappa \epsilon^2  \mathcal{D}[\hat{c}_a^\dag \hat{c}_s]+ \mathcal{L}_\epsilon.
\end{equation}
By comparison of
Eqs.~(\ref{eq:EffCavityME}) and (\ref{Lth}) we identify $\kappa'=\epsilon^2
\kappa$, with the parameter $\epsilon^2\equiv (\Delta/\delta)^2$ determining the minimal
effective temperature of the engineered two-photon process. For typical
parameters, $\omega_{\rm q}=10$ GHz, $\delta=1$ GHz and $\Delta_e=0.8$ GHz, we obtain
$\epsilon=0.2$, and hence the limit $\kappa'\rightarrow 0$ considered in most
parts of the paper can be implemented to a very good approximation. 
However, we point out that the ratio $\kappa'/\kappa$ can always be increased artificially, for example, by adding an additional coherent or incoherent driving field to populate the
state $|S\rangle$.  Our derivation also shows the existence of coherences between $\hat{c}_s^\dag \hat{c}_a$ and $\hat{c}_a^\dag \hat{c}_s$, which lead to additional squeezing terms of order $\mathcal{O}(\kappa \epsilon)$, and that in Eq.~(\ref{expandLth}) are summarized by $ \mathcal{L}_\epsilon$.  
For $\epsilon \ll1 $, and in the presence of cavity losses, we expect the influence of these coherences to be negligible, while the relevant effect on the populations is already captured by a finite $\kappa'$. Indeed, numerical simulations similar to those discussed in Sec.~\ref{Results} show no significant differences between the exact ME~(\ref{eq:EffCavityME}) and our model Liouvillian~(\ref{Lth}), even up to values of $\epsilon\approx 0.5$. However, by choosing $\epsilon\approx 1$ and $\Gamma\ll \kappa$, it would be possible to extend our model and study physical effects directly linked to the parameter $\epsilon$, such as number-conserving squeezing as previously analyzed in \cite{WM2011}.

\begin{figure}
\begin{center}
\includegraphics[width=0.48\linewidth]{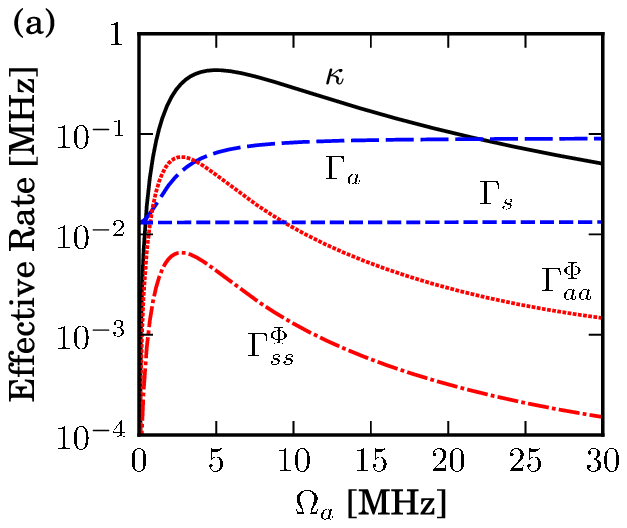}
\includegraphics[width=0.48\linewidth]{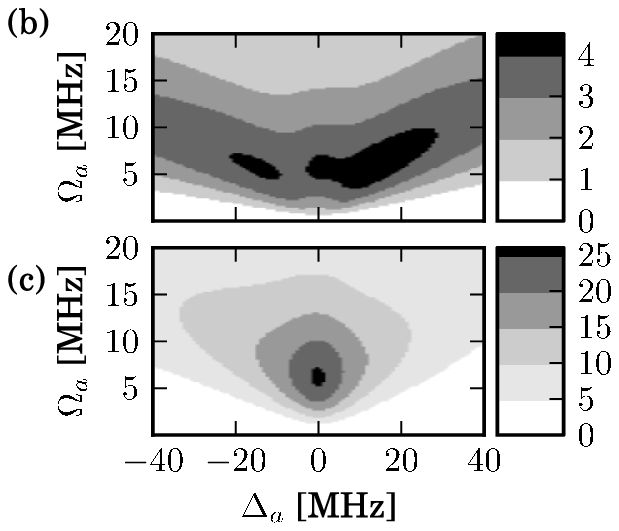}
\caption{\label{FOM} a) Single and two-photon rates which appear in the
  effective model of Eq.~(\ref{eq:EffCavityME}).  The photon scattering rate of
  interest, $\kappa$, becomes significantly larger than the others in a certain
  region of the parameter space.  Here we have chosen $g=25$ MHz, $\delta=1$
  GHz, $\Delta_s=\Delta_a=0$ MHz, $\Delta_e=800$ MHz, $\gamma=10$ MHz, $
  \Gamma =10$ kHz. b) Figure of merit for the validity of our model for high-momentum 
  modes, $F_a\equiv\kappa/(\Gamma_a+\Gamma_{aa}^\Phi)$, 
  as a function of qubit driving $\Omega_a$ and detuning $\Delta_a$. Rest of parameters as in (a).  
  c) The same plot as in (b) but for low-momentum modes, $F_s\equiv\kappa/(\Gamma_s+\Gamma_{ss}^\Phi)$. }
\end{center}
\end{figure}

Apart from the two-photon scattering processes of interest, the coupling to the
qubits introduces various imperfections.  On the one hand, these are the
enhanced cavity decay rates $\Gamma_{s,a}$ introduced above, and on the other hand,
we obtain additional dephasing terms $\Gamma_{ij}^\Phi$ due to fluctuating Stark
shifts when the qubit jumps incoherently between states $|G\rangle$ and
$|A\rangle$.  Since the ratio $\kappa/\Gamma_{s,a}$ scales as $\sim
g^2/\gamma^2$, single photon losses can be suppressed in the strong coupling
regime, as long as $\kappa$ also exceeds the bare decay rate $ \Gamma $.  The
ratio $\kappa/\Gamma_{ij}^\Phi$ is independent of $g$, and depends on the details
of the correlation functions (\ref{spectrumEq}) and therefore on the parameters
$\Omega_a$, $\Delta_{s,a}$.  In Fig.~\ref{FOM}a, $\kappa$ is compared with the
other decoherence rates, and we see that an experimentally feasible parameter
range exists, where $\kappa$ is the dominant rate. The main
corrections arise from the enhanced decay $\Gamma_a$ and the dephasing
$\Gamma_{aa}^\Phi$ of the antisymmetric mode, while the corresponding rates for
the symmetric mode are much smaller.  In Fig.~\ref{FOM}b and \ref{FOM}c we plot
the two quantities $F_{\eta}\equiv \kappa / (\Gamma_\eta +
\Gamma_{\eta\eta}^\Phi)$, $\eta=s,a$, as a function of the control parameters $\Omega_a$, $\Delta_a$. 
These quantities represent a simple figure of merit for characterizing the validity of
our model for low ($F_s$) and high ($F_a$) momentum modes. We see that within a
large parameter regime, the number conserving photon scattering rate $\kappa$ can
exceed all decoherence processes discussed here, and therefore the proposed model in Eq.~(\ref{effectivecavity}) 
provides a reliable description of the system
dynamics. In particular, this is true for the low momentum (symmetric) modes, where condensation occurs. 

Finally, let us briefly comment on the effective photon-photon interactions
described by $\hat{\cal H}_{\rm eff}$.  For a resonant two-photon process,
$\Delta_s=0$, we find $U_{\rm eff}\approx 0$.  Moreover,
$U_{\eta\eta'}^\Phi\approx \Gamma_{\eta\eta'}^\Phi < \kappa$ for the typical
parameters considered above, and in particular these interactions vanish for
$\Delta_a=0$.  Therefore, in this work we focus exclusively on the purely
dissipative cavity dynamics.  However, by setting $|\Delta_s|>\gamma$ instead,
the regime of strong coherent photon-photon coupling $U_{\rm
  eff}>\kappa>\Gamma_{\eta\eta'}^\Phi,\Gamma_{s,a}$ can be engineered as well,
having $\hat{\cal H}_{\rm eff}\simeq U_{\rm eff} \hat{c}_s^\dag \hat{c}_s
\hat{c}_a^\dag \hat{c}_a$.

In summary, we find that the ME~(\ref{effectivecavity}) can be
implemented with superconducting microwave cavities under realistic experimental
conditions.  Our analysis, which has been detailed here for generic coupled two level
systems, can be thus adapted to various nonlinear Josephson
devices, where the parameter $\lambda_{x,z}$ plays the role of the nonlinearity.

\section{Condensation of photons: two cavities} \label{Results}

\subsection{Semiclassical approximation}

In the previous section we have shown how to implement the effective ME
(\ref{effectivecavity}). Now, we discuss how the scattering between photons
induced by the Liouvillian ${\cal L}_{\kappa}$ leads to condensation of
photons in an open and driven cavity array. In this section we first illustrate
the basic process for the minimal case with $L=2$ cavities only. As already
mentioned above, we assume $U=0$, and consider the case where only the
antisymmetric mode is excited, i.e. $\Omega_1=-\Omega_2=\Omega/\sqrt{2}$.  In the rotating frame with respect to
the drive frequency $\omega_{\rm d}$ the
equations of motion (EOM) for the field expectation values
$\langle\hat{c}_{s}\rangle$ and $\langle\hat{c}_{a}\rangle$ of the symmetric and
antisymmetric cavity modes are
\begin{equation}
\begin{array}{l} \label{EOMcoherentpumping}
\partial_{t} \langle \hat{c}_s \rangle = [- i (\delta_{\rm d}-2J) - \Gamma] \langle \hat{c}_s \rangle + \kappa \langle \hat{c}_s \hat{c}_a^{\dag} \hat{c}_a \rangle,\\
\partial_{t} \langle \hat{c}_a \rangle = [- i \delta_{\rm d} - \Gamma ] \langle \hat{c}_a \rangle - i \Omega - \kappa \langle (\hat{c}_s^{\dag} \hat{c}_s + 1) \hat{c}_a \rangle,
\end{array}
\end{equation}
where $\delta_{\rm d}=\omega_{\rm c}+J-\omega_{\rm d}$ is the detuning of the
antisymmetric mode with respect to the driving field. For the moment, we assume
$\kappa' = 0$ in Eq.~(\ref{Lth}) and postpone the discussion of this term to the end of the
present section.  The EOM for the field expectation values couple to third-order
correlation functions, starting a hierarchy of equations that cannot be solved
analytically.  To truncate the hierarchy, we resort here to a semiclassical
approximation in which the state of the system is assumed to be coherent.  With
this assumption, each normal-ordered correlation function can be readily
substituted by the corresponding product of field expectation values.  The EOM
read
\begin{equation}
\begin{array}{l} \label{meanfieldEOM2cavities}
\partial_{t} s = - i (\delta_{\rm d}-2J) s - (\Gamma - \kappa |a|^{2})s, \\
\partial_{t} a = - i \delta_{\rm d} a - i  \Omega - [\Gamma + \kappa(|s|^{2} + 1)] a,
\end{array}
\end{equation}
where $s \equiv \langle \hat{c}_s \rangle$ and $a \equiv \langle \hat{c}_a
\rangle$ for brevity.  The semiclassical approximation, in general, is valid for
large photon numbers but it offers a satisfactory explanation of the dynamics of
the system for smaller occupation as well, as we verify below with the numerical
integration of the EOM~(\ref{EOMcoherentpumping}).  

\begin{figure}
\begin{center}
\includegraphics[width=0.48\linewidth]{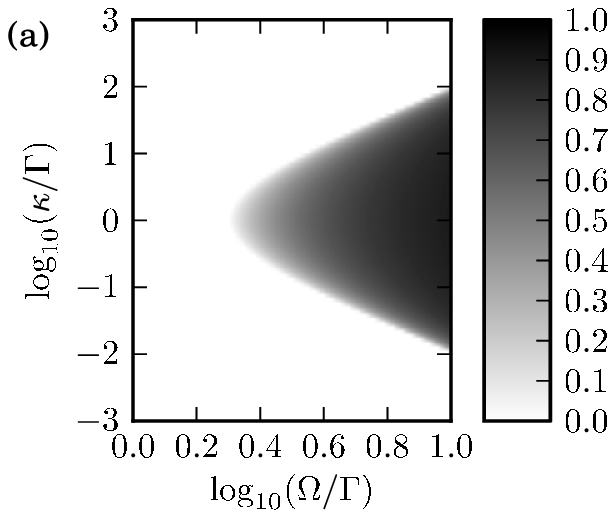}
\includegraphics[width=0.48\linewidth]{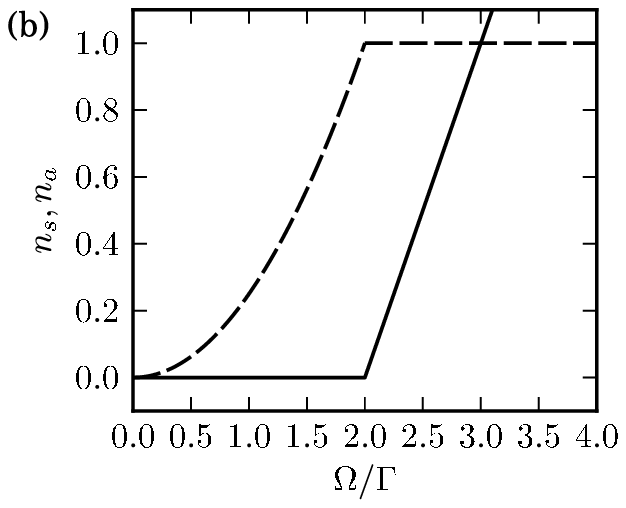}
\caption{\label{transitionmeanfield} Results (\ref{semiclassicalresults}) of the
  semiclassical approximation to the EOM~(\ref{EOMcoherentpumping}). a)
  Population fraction $n_s/n$ in the symmetric mode as a function of cavity
  pumping $\Omega$ and dissipation strength $\kappa$. b) Population in the
  symmetric (solid line) and antisymmetric (dashed line) mode for
  $\kappa=\Gamma$ as a function of the pumping strength $\Omega$.  The threshold
  (\ref{threshold}) is $\Omega_{\rm crit} = 2.0 \, \Gamma$. }
\end{center}
\end{figure}

The exact equations (\ref{EOMcoherentpumping}) become nonlinear $c$-number equations for the expectation
values $s$ and $a$.  Nonlinear equations may have multiple, stable and unstable
steady-state solutions.  A paradigmatic example of this behavior are the
classical equations of the laser~\cite{HakenBook}, where a solution which lacks
coherent emission is possible at any pumping strength, but is unstable above a
certain threshold.  Here we find a comparable result, with a threshold value
\begin{equation}\label{threshold}
\Omega_{\rm crit} \equiv \Gamma\sqrt{\frac{\delta_{\rm d}^2 + (\Gamma+\kappa)^2}{\Gamma\kappa}},
\end{equation}
for the pumping strength $\Omega$.  Below the threshold, the only stable
steady-state solution to Eq.~(\ref{meanfieldEOM2cavities}) is (c.f. also
\ref{AppendixB})
\begin{eqnarray} \label{solbelowthreshold}
|s|^2=0, \qquad |a|^2= \frac{\Omega^2}{\delta_{\rm d}^2+(\Gamma+\kappa)^2},
\end{eqnarray}
and only the pumped antisymmetric mode is populated. For driving strengths
above the threshold we obtain the stable solution
\begin{eqnarray} \label{solabovethreshold}
|s|^2=\sqrt{\frac{\Omega^2}{\Gamma\kappa} -\frac{\delta_{\rm d}^2}{\kappa^2}} - \frac{\Gamma}{\kappa} - 1, \;\;\;\;\; |a|^2= \frac{\Omega^2}{\delta_{\rm d}^2+[\Gamma+\kappa(|s|^2+1)]^2},
\end{eqnarray}
where the symmetric mode features nonvanishing occupation. 
Note that these equations depend on $J$ only via the
detuning $\delta_{\rm d}$, which has the only effect to reduce the
driving strength. Therefore, from now on we will restrict ourselves to
$\delta_{\rm d}=0$ and $J\rightarrow 0$. The stationary solution to the
semiclassical EOM then simplifies to
\begin{equation}
\begin{array}{l} \label{semiclassicalresults}
|s|^2 = 0, \qquad |a|^2=\frac{\Omega^2}{(\Gamma+\kappa)^2}, \qquad \mbox{if} \quad \Omega < \Omega_{\rm crit}, \\
|s|^2 = \frac{\Omega}{\sqrt{\Gamma\kappa}} - \frac{\Gamma}{\kappa} - 1, \qquad |a|^2=\frac{\Gamma}{\kappa}, \qquad \mbox{if} \quad \Omega > \Omega_{\rm crit}.
\end{array}
\end{equation}
The behavior of the populations $n_s\equiv |s|^2$ and $n_a\equiv |a|^2$ is
plotted in Fig.~\ref{transitionmeanfield} as a function of the pumping intensity
and the effective decay $\kappa$.  Below threshold, the symmetric mode is empty
and the antisymmetric mode occupation grows quadratically with the pumping
intensity.  For $\Omega>\Omega_{\rm crit}$, the antisymmetric mode population
saturates, and the symmetric mode occupation grows linearly.  Thus, if $\Omega$
is sufficiently large, most of the photon population is transferred to the
symmetric mode.  This transition to a state in which photons `macroscopically'
occupy the symmetric superposition of two cavity modes is a direct consequence
of the engineered nonlinear Liouvillian~(\ref{Lth}).

The discussion of Eqs.~(\ref{semiclassicalresults}) illustrates both, similarities and differences between the
present setup and a laser \cite{AINYPNT2007,RIA2007,YLSN2007,HFHSS2008}.  In a single- or multi-mode laser, the light field is coupled to a pumped reservoir and the lasing threshold corresponds to the point where the total light intensity 
starts to grow.  In our case, since the antisymmetric mode is driven directly by a classical field,  the total photon population is always finite and grows monotonically.  The threshold then marks the point where the many-body scattering mechanism contained in ${\cal L}_{\kappa}$ dominates, and a macroscopic transfer of photons from  the antisymmetric to the symmetric mode occurs. In this sense, one can rather think of the antisymmetric mode as a photon reservoir from which a condensation of photons (`lasing') into the symmetric mode occurs. However, as we will see below, this analogy between lasing and condensation \cite{Snoke2003,SL2002,SLS2003,Scully1999,KSZZ2000} holds in our system only for certain parameter regimes, and in particular striking differences can be observed in the regime of low photon numbers.     

Finally, note that both for $\kappa\to 0$ as well as for $\kappa \rightarrow \infty$
the condensation is suppressed, and that the phase diagram in
Fig.~\ref{transitionmeanfield} is symmetric with respect to the ratio
$\kappa/\Gamma$.  For small $\kappa$, not enough photons are scattered into the
symmetric mode, while for large $\kappa$ the antisymmetric mode is highly
overdamped and the population of the whole system is small as in the first place.

\subsection{Exact diagonalization}

\begin{figure}
\begin{center}
\includegraphics[width=0.48\linewidth]{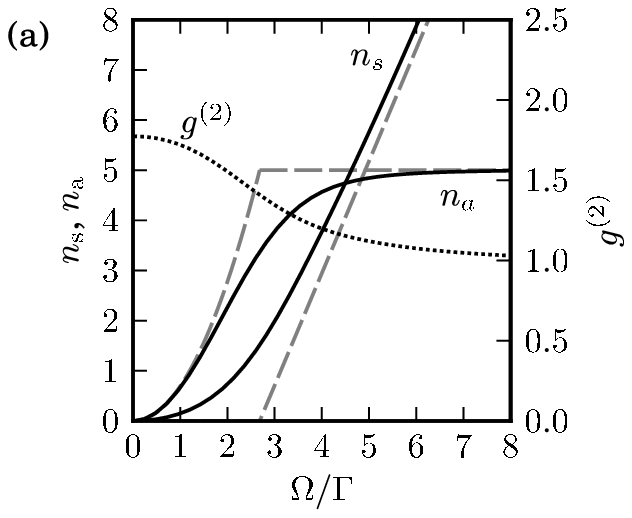}
\includegraphics[width=0.48\linewidth]{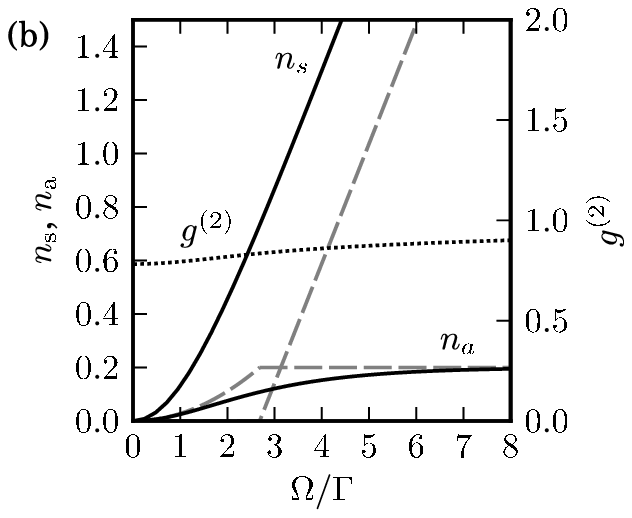}
\caption{\label{numerical2cavities} Numerical solution to the exact equations of
  motion for $L=2$ cavities.  The occupation numbers $n_s$ and $n_a$ of the symmetric and antisymmetric mode
are plotted as a function of the coherent driving strength $\Omega$ for a) $\kappa/\Gamma=0.2$ and b) $\kappa/\Gamma=5$. In both plots $\delta_{\rm d}=0$, and the dashed lines indicate the corresponding semiclassical predictions given by Eq.~(\ref{semiclassicalresults}).   
The dotted lines show the two-photon correlation function of the symmetric mode $g^{(2)}=\langle
\hat{n}_s (\hat{n}_s - 1) \rangle / \langle \hat{n}_s \rangle^{2}$.
}
\end{center}
\end{figure}

To support the results of the semiclassical approximation discussed above, we
resort to the exact diagonalization of the system composed by two cavities.  
The time-evolution of the system for typical parameters features an
initial transient in which the photonic population accumulates in the
antisymmetric mode, which is directly pumped, and a later stage in which
photons scatter into the symmetric mode.  Finally, a steady state is reached in
which a dynamical equilibrium takes place between the populations in the two
modes.  We remark that the population in the modes is continuously subjected to
losses.

The steady-state values of the populations $n_{s}=\langle \hat c_s^\dag \hat c_s\rangle $ and $n_{a}=\langle \hat c_a^\dag \hat c_a\rangle$ in the symmetric
and antisymmetric mode, respectively, are shown in
Fig.~\ref{numerical2cavities} as a function of the pumping strength $\Omega$ and for two different values of $\kappa/\Gamma$. 
In the limit $\kappa\ll\Gamma$, where we expect the transition to occur at large mode occupation numbers $n_a$, we find that the exact solution matches qualitatively and quantitatively very well the semiclassical predictions. In particular, we see that above $\Omega_{\rm crit}$ the population $n_{s}$ grows linearly with the driving strength, while the antisymmetric mode saturates at a value $n_a=\Gamma/\kappa$.  Differently from the prediction
of the semiclassical analysis, however, $n_{s}$ does not vanish exactly below
the threshold but it is smaller than $n_{a}$. A study of the equal-time two-photon correlation $g^{(2)} \equiv \langle
\hat{n}_s (\hat{n}_s - 1) \rangle / \langle \hat{n}_s \rangle^{2}$ reveals that, when crossing $\Omega_{\rm crit}$, the photon statistics of the symmetric mode changes from a thermal ($g^{(2)}\approx 2$) to a Poissonian ($g^{(2)}\approx 1$) distribution. This aspect is in agreement with the standard lasing transition. 

In the opposite regime, $\kappa\gg\Gamma$,  Eq.~(\ref{semiclassicalresults}) predicts that the population $n_a$ of the driven mode is always less than one, and we do not expect the semiclassical analysis to give accurate results. Indeed, Fig.~\ref{numerical2cavities}b shows that in this limit the transition is completely washed out and $n_s$ exceeds $n_a$ for all pumping strengths. Nevertheless,  above $\Omega_{\rm crit}$ we still observe the characteristic saturation of $n_a$ and  -- apart from an additional offset -- the correct linear scaling of the population in the symmetric mode. However, in strong contrast to the semiclassical regime, we find that the photon statistics of the symmetric mode is sub-Poissonian ($g^{(2)}< 1$) at low driving strengths and approaches the classical limit $g^{(2)}= 1$ from below. This anti-bunching effect can be understood from the fact that the effective damping of the $\hat c_a$ mode is given by $\Gamma +\kappa(\langle \hat n_s\rangle +1)$. Therefore, the scattering of the first photon into the symmetric mode changes the damping significantly and suppresses the following repopulation of the `reservoir mode' $\hat c_a$. The limit $\kappa\gg \Gamma$ and $\Omega\approx \Omega_{\rm crit}$  would then result in a `single photon condensate' with $ n_a\ll n_s \sim 1$ and a non-classical statistics with $g^{(2)}\rightarrow 0.5$.

\subsection{Effective temperature}

Finally, let us briefly comment on the effect of the reverse photon scattering
process $\sim \kappa'$ in Eq.~(\ref{Lth}). As we have discussed in
Sec.~\ref{cavityarrayintro}, in the case of two isolated cavities the
presence of the reverse photon scattering term can be interpreted as a finite
temperature effect, and we can set $\kappa'=\kappa e^{-2 J/k_BT_{\rm
    eff}}$. By including this term, the EOM are
\begin{equation}
\begin{array}{l}
\partial_{t} \langle \hat{c}_s \rangle = - (\Gamma+\kappa') \langle \hat{c}_s \rangle + (\kappa-\kappa') \langle \hat{c}_s \hat{c}_a^{\dag} \hat{c}_a \rangle, \\
\partial_{t} \langle \hat{c}_a \rangle = - (\Gamma+\kappa') \langle \hat{c}_a \rangle - i  \Omega - (\kappa-\kappa') \langle (\hat{c}_s^{\dag} \hat{c}_s + 1) \hat{c}_a \rangle.
\end{array}
\end{equation}
These equations are of the same form of (\ref{EOMcoherentpumping}), with the
replacements $\Gamma\to\Gamma+\kappa'$ and $\kappa\to\kappa-\kappa'$.
Therefore, in the semiclassical regime we obtain the same physics, but with a
renormalized pumping threshold
\begin{equation}\label{finiteTthreshold}
\Omega_{\rm crit} (T_{\rm eff})\equiv (\Gamma+\kappa) \sqrt{\frac{(\Gamma/\kappa + e^{-2 J/k_BT_{\rm eff}}  )}{(1-e^{-2 J/k_BT_{\rm eff}} )}} .
\end{equation}
In particular, for any fixed driving strength we
can use Eq.~(\ref{finiteTthreshold}) to define a critical effective temperature
above which photon condensation does not occur.

\section{Photon condensation in a cavity array} \label{cavityarraysec}

We turn now to the solution of Eq.~(\ref{effectivecavity}) in a lattice with $L
\gg 1$ cavities.  The central goal of this section is to show that the many-body
Liouvillian (\ref{Lth}) induces condensation of photons into the homogeneous
lattice mode with zero momentum $q$.  Following the scheme introduced in the
previous section in the case of $L = 2$ cavities only, we consider the scenario in
which each cavity mode $\hat{c}_{\ell}$ in the array, with $\ell = 1, \dots L$,
is coherently driven by a classical field with staggered amplitude
$\Omega_{\ell} = (-1)^{\ell} \Omega$.  In this way, the coherent driving acts on
the edge of the Brillouin zone, while dissipation-induced condensation is expected to take place at
the center.

For simplicity, we assume periodic boundary conditions for the lattice.  We
rewrite Eq.~(\ref{effectivecavity}) in terms of the annihilation operators $\hat{c}_q = \frac{1}{\sqrt{L}} \sum_\ell e^{i q \ell} \hat{c}_\ell$ of the photonic modes in momentum space, 
where $q$ takes values $-\pi (L-2)/L,\dots,\pi
(L-2)/L,\pi$ (given here in units of the inverse lattice spacing) in the discretized Brillouin zone. In the
rotating frame with respect to the coherent driving, the Hamiltonian is then given by
\begin{equation} \label{Hamiltonianq}
\hat{\cal H}_{\rm c} + \hat{\cal H}_{\Omega} = \sum_{q} [\delta_{\rm d}-2J(1+\cos(q))] \hat{c}_{q}^{\dag} \hat{c}_{q} + \sum_{q} ( \Omega_{q} \hat{c}_{q}^{\dag} + \Omega_{q}^{\ast} \hat{c}_{q}),
\end{equation}
where $\Omega_q
= \frac{1}{\sqrt{L}} \sum_{\ell} e^{iq\ell}\Omega_{\ell}$ and $\delta_{\rm d}= \omega_{\rm c}+2J-\omega_{\rm d}$ is the detuning of the $q=\pi$ mode from the driving frequency.
The Liouville operators are
\begin{equation} \label{LiouvillianGammaq}
\Gamma \sum_{\ell} {\cal D}[\hat{c}_{\ell}] \rho = \Gamma \sum_{q} [2 \hat{c}_{q} \rho \hat{c}_{q}^{\dag} - \hat{c}_{q}^{\dag} \hat{c}_{q} \rho - \rho \hat{c}_{q}^{\dag} \hat{c}_{q}],
\end{equation}
and
\begin{eqnarray} \label{Liouvilliankappaq}
{\cal L}_{\kappa} \rho & = & \sum_{q_{1}q_{2}q_{3}q_{4}} \delta(q_{1}+q_{3}-q_{2}-q_{4}) K(q_{1},q_{2},q_{3},q_{4}) \nonumber \\
& & \times [2 \hat{c}_{q_{1}}^{\dag} \hat{c}_{q_{2}} \rho \hat{c}_{q_{3}}^{\dag} \hat{c}_{q_{4}} - \hat{c}_{q_{3}}^{\dag} \hat{c}_{q_{4}} \hat{c}_{q_{1}}^{\dag} \hat{c}_{q_{2}} \rho - \rho \hat{c}_{q_{3}}^{\dag} \hat{c}_{q_{4}} \hat{c}_{q_{1}}^{\dag} \hat{c}_{q_{2}}].
\end{eqnarray}
In the latter equation, the function $K$ plays the role of a scattering kernel,
which has the general form
\begin{eqnarray}
K(q_{1},q_{2},q_{3},q_{4}) =  \frac{4}{L} \Big[ &&\kappa \cos\frac{q_{1}}{2} \sin\frac{q_{2}}{2} \sin\frac{q_{3}}{2} \cos\frac{q_{4}}{2} \nonumber \\
&&+\kappa'\sin\frac{q_{1}}{2} \cos\frac{q_{2}}{2} \cos\frac{q_{3}}{2} \sin\frac{q_{4}}{2}\Big].
\end{eqnarray}
In the following analysis, we will focus on the case $\kappa'=0$. Furthermore, as
discussed above, the photon condensation effect does not depend on the tunneling
amplitude and we can assume $J \rightarrow 0$ in the Hamiltonian.

Solving the ME~(\ref{effectivecavity}) in an extended array is in general a very
demanding task,
and exact diagonalization can only be provided for very limited
system size (see our results in Fig.~\ref{condensation_fintite_q}b for $L=4$
cavities).
For this reason, we resort in the following to two complementary approximations, which
allow us to treat the full dynamics of the photonic population in the whole
Brillouin zone.  First we consider the equation of motion for the field
expectation values within the semiclassical approximation, which was discussed
in Sec.~\ref{Results} for the case $L = 2$, and showed reliable results for large photon numbers.
Then we treat the photonic population directly and derive a Boltzmann-like
equation starting from the exact EOM for the two-point correlation functions.
Within the latter approximation we are able to estimate the finite correlation
length of the photonic field in the array.
In this respect, this technique fills the gap
between the semiclassical analysis of an infinitely extended array and the
numerical solution to a small-size system. We emphasize that all techniques agree on the central results that we present here, i.e.~the existence of a
macroscopic photonic population at the center of the Brillouin zone.

\subsection{Semiclassical approximation}

\begin{figure}
\begin{center}
\includegraphics[width=0.48\linewidth]{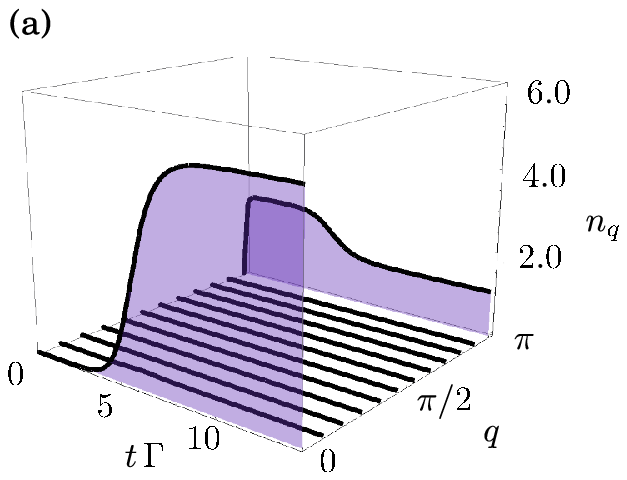}
\includegraphics[width=0.48\linewidth]{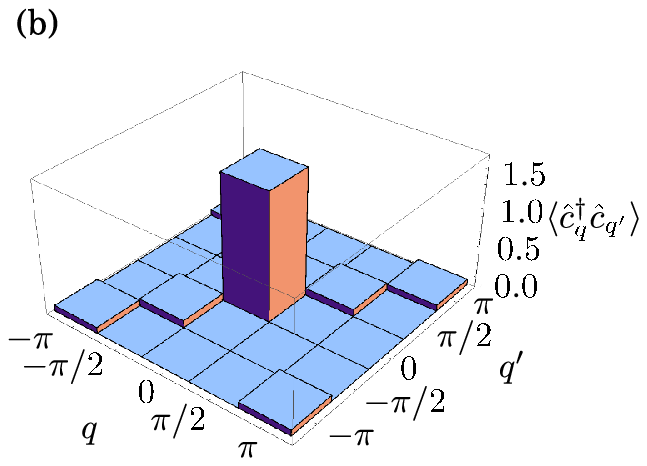}
\caption{\label{condensation_fintite_q} a) Numerical solution of the
  semiclassical equations of motion (\ref{eq:boltzmanncoherent}). $L=22$
  momentum states are used in the integration and the population $n_{q} =
  |\psi_{q}|^{2}$ is shown in the positive half $q>0$ of the Brillouin zone.  b)
  One-particle density matrix $\langle \hat{c}^{\dag}_{q} \hat{c}_{q'} \rangle$
  in momentum space, obtained with the numerical integration of the exact
  equations of motion for $L=4$ cavities, integrated until $t_{\rm max} \,
  \Gamma = 0.5$.  In both panels we use $\kappa = 5.0 \, \Gamma$ and
  $\Omega_{q=\pi} = 3.5 \sqrt{L} \, \Gamma$. }
\end{center}
\end{figure}

For many cavities, the EOM for the cavity fields $\psi_{q} = \langle \hat{c}_{q} \rangle$
(analogous to Eq.~(\ref{meanfieldEOM2cavities})) read
\begin{eqnarray}\label{eq:boltzmanncoherent}
\partial_{t} \psi_{q} & = & [- i \delta_{\rm d} - \Gamma - \kappa (1 - \cos{q})] \psi_{q} - i \Omega_{q} \nonumber \\
& & + \frac{ \kappa}{L} \sum_{q_{1}q_{2}} \psi_{q_{1}} \psi_{q_{2}} \psi_{q_{1} + q_{2} - q}^{\ast}
[\cos{q} - \cos{(q_{1} + q_{2} - q)}],
\end{eqnarray}
where $\Omega_{q} = \delta_{q,\pi} \Omega_\pi $.  
The numerical solution of this system of nonlinear coupled equations is
straightforward. Fig.~\ref{condensation_fintite_q}a shows the population $n_{q} = |\psi_{q}|^{2}$ on the
discretized Brillouin zone for $L = 22$ cavities.  We see that, in the initial
stage of the dynamics, the coherently pumped mode at $q = \pi$ is populated on
the time scale of $\Gamma^{-1}$.  Later, photonic population is transferred to
the mode $q = 0$ and the steady state is reached.  A surprising feature is that
the population of the intermediate modes in the Brillouin zone is almost
vanishing (and indeed not visible in Fig.~\ref{condensation_fintite_q}a)
throughout the time-evolution.  This contrasts sharply with typical relaxation
phenomena in the Brillouin zone (see e.g.~Ref.~\cite{HaugJauhoBook}) where the
excited population is expected to drift to lower energy states by
interaction-induced relaxation.  The reason for this behavior is
the specific form of the engineered system-bath coupling ${\cal L}_{\kappa}$.

We can take advantage of the negligible population in the intermediate modes of
the Brillouin zone and introduce a two-mode approximation, in which
only the averages $\psi_{0}$ and $\psi_{\pi}$ of the modes $q = 0$ and $q =
\pi$, respectively, are assumed finite. The EOM (\ref{eq:boltzmanncoherent}) then
reduce to
\begin{equation}
\begin{array}{l}
\partial_{t} \psi_{0} =  - (\Gamma - 4 \kappa ~ \frac{1}{L} |\psi_{\pi}|^{2}) \psi_{0}, \\
\partial_{t} \psi_{\pi} = - i \Omega_{\pi} - [ \Gamma + 4 \kappa ~ (\frac{1}{L}|\psi_{0}|^{2} + 1/2) ] \psi_{\pi}.
\end{array}
\end{equation}
The structure of these equations is exactly the same as Eq.~(\ref{meanfieldEOM2cavities}), with the $|a|^2/2$ and $|s|^2/2$ replaced by the photon densities $|\psi_\pi|^2/L$ and  $|\psi_0|^2/L$, respectively. The analytical discussion of these coupled equations follows the
same path as that from Eq.~(\ref{meanfieldEOM2cavities}) to
Eq.~(\ref{semiclassicalresults}).
The threshold, above which photons start to condense into the $q=0$ mode, is
$\Omega_{\pi,{\rm crit}}= \sqrt{L/2}\times \sqrt{\Gamma (\Gamma + 2 \kappa)^2 /
  2 \kappa}$. Note that compared to the results presented in the previous section, a factor $2\kappa$, instead of $\kappa$, appears. This is simply due to the fact that in a 1D array (with periodic boundary conditions) each cavity is dissipatively coupled to two neighboring sites.    

In Fig.~\ref{condensation_fintite_q}b we also present the results of the exact
diagonalization of a small cluster of $L = 4$ cavities, for typical parameters
comparable to the semiclassical results.  We show the absolute values of the
entries of the one-particle density matrix in momentum space, $\langle
\hat{c}_{q}^{\dag} \hat{c}_{q'} \rangle$.  The values along the diagonal are the
populations $n_{q}$, and the peak at $q = 0$ clearly demonstrates substantial
occupation of that mode.  The non-diagonal entries, which represent the
correlation between modes with different momentum, are small.  These values
cannot be obtained correctly from the semiclassical approximation, where
correlations factorize and one would have $|\langle \hat{c}_{q}^{\dag} \hat{c}_{q'}
\rangle| \simeq \sqrt{n_{q}} \sqrt{n_{q'}}$.  The absence of correlation between
modes with different momentum can be understood, since the photonic scattering
term originates from the non-unitary contribution to the ME
(\ref{effectivecavity}).  This fact suggests that an approximation scheme should
hold, where the correlations between momentum modes are neglected from the
beginning, and only the populations $n_{q}$ are the dynamical variables.  The
Boltzmann-like EOM that implements this scheme is presented below.

\subsection{Steady-state photon distribution}

\begin{figure}
\begin{center}
\includegraphics[width=0.48\linewidth]{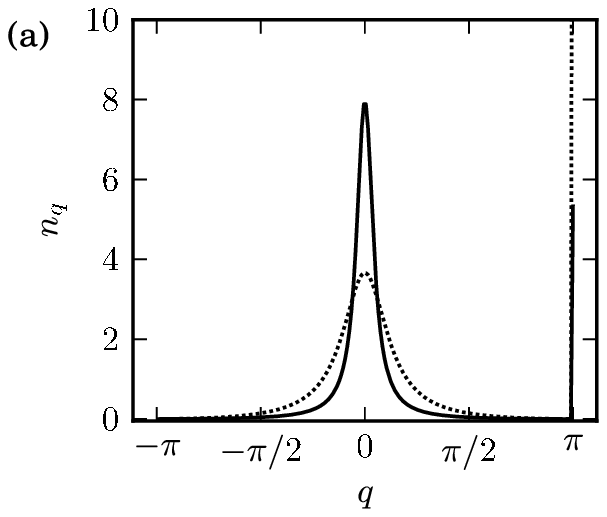}
\includegraphics[width=0.48\linewidth]{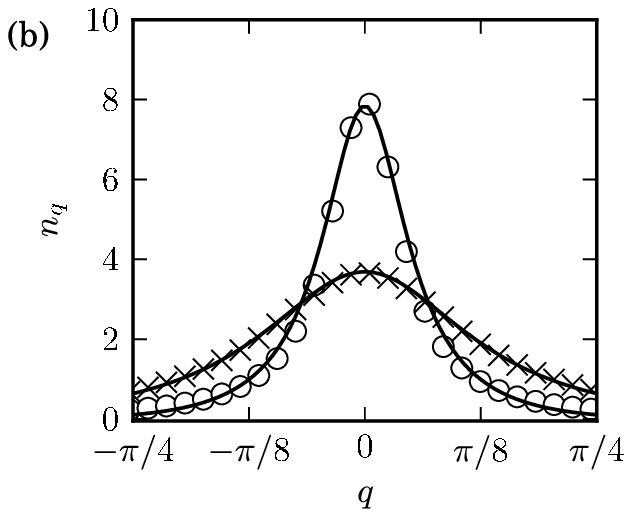}
\caption{\label{figurenq} Results of the numerical integration of the
  Boltzmann-like equations for the populations $n_{q}$ on a discretized
  Brillouin zone with $L = 201$ points, until $t_{\rm max} \, \Gamma = 10.0$,
  for $\kappa = 5.0 \, \Gamma$, $\Omega_{\pi} = 3.5 \sqrt{L}\Gamma$ (solid line in (a)
  and circles in (b)) and $\kappa = 1.0$, $\Omega_{\pi} = 2.0 \sqrt{L}\Gamma$ (dotted
  line in (a) and crosses in (b)).  Two Bose-Einstein equilibrium distributions
  are plotted in (b) (solid lines) for comparison. }
\end{center}
\end{figure}

The exact EOM for the populations $n_{q}=\langle
\hat{c}_{q}^{\dag} \hat{c}_{q} \rangle$ are
\begin{eqnarray}\label{cqdagcq}
\partial_{t} \langle \hat{c}_{q}^{\dag} \hat{c}_{q} \rangle &=& -2 [ \Gamma + \kappa (1 - \cos{q})] \langle \hat{c}^{\dag}_{q} \hat{c}_{q} \rangle + i \Omega_{q}^{\ast} \langle \hat{c}_{q} \rangle - i \Omega_{q} \langle \hat{c}_{q}^{\dag} \rangle \nonumber\\ &+& \frac{8 \kappa}{L}  \sum_{q_1} \cos^2\frac{q}{2}  \sin^2\frac{q_1}{2} \langle \hat{c}^{\dag}_{q_1} \hat{c}_{q_1} \rangle + {\cal S}(q),
\end{eqnarray}
where the term ${\cal S}(q)$ contains fourth-order correlation functions, which
are not diagonal in momentum space, and reads
\begin{equation}
\mathcal{S}(q)=   \frac{\kappa}{L} \sum_{q_{1}q_{2}}Ê \left[\cos(q)-\cos(q_1+q_2-q)\right]  \langle c_{q_2+q_1-q}^\dag  c_{q}^\dag c_{q_1}  c_{q_2} \rangle +  {\rm H.c.} 
\end{equation}
One route to reduce the EOM to a manageable form is to implement a truncation in
the infinite hierarchy of coupled correlation functions, expressing higher-order
correlations in terms of products of lower-order ones.  Here, we reduce the
fourth-order correlation function to a product of second-order diagonal
correlation functions according to the prescription
\begin{equation} \label{HFfactorization}
\langle \hat{c}^{\dag}_{q_{1}} \hat{c}^{\dag}_{q_{2}} \hat{c}_{q_{3}} \hat{c}_{q_{4}} \rangle
\simeq (\delta_{q_{1},q_{4}}\delta_{q_{2},q_{3}} + \delta_{q_{1},q_{3}}\delta_{q_{2},q_{4}}) \langle \hat{c}^{\dag}_{q_{1}} \hat{c}_{q_{1}} \rangle \langle \hat{c}^{\dag}_{q_{2}} \hat{c}_{q_{2}} \rangle.
\end{equation}
This prescription is in fact a Hartree-Fock
approximation~\cite{FetterWaleckaBook}, because we assume that the dominant
contribution to the correlations arises in the density channel of the momentum
modes.  The motivations for this choice have been discussed above.  Using this
factorization we obtain the following simplified form for the scattering term
\begin{equation}
\mathcal{S}(q)\simeq    \frac{4 \kappa}{L} \sum_{q_{1}}Ê \left[\cos(q)-\cos(q_1)\right]  \langle c_{q_1}^\dag  c_{q_1}\rangle \langle c_q^\dag c_{q}  \rangle.
\end{equation}
A self-consistent solution is to assume that $\langle \hat{c}_{q} \rangle = 0$
for $q \neq \pi$.  The EOM~(\ref{cqdagcq}) become then a set of coupled,
nonlinear, Boltzmann-like equations~\cite{HaugJauhoBook} for the populations only.  However, on the
edge of the Brillouin zone, the coherent pumping $\Omega_{\pi}$ couples to the
average field, which does not vanish.  The numerical integration of the
resulting equations is straightforward and produces the distribution $n_{q}$ in
the whole Brillouin zone.  A finer discretization in the Brillouin zone is
possible, with respect to the semiclassical equations
(\ref{eq:boltzmanncoherent}), with a comparable numerical effort.  Typical
profiles of the steady-state distributions are shown in Fig.~\ref{figurenq}.
Due to the increased resolution in momentum space, the broadening of the
condensate peak at $q = 0$ can be appreciated, and its dependence on the
parameters is discussed below.

To discuss the width of the central peak, which is related to the correlation
length $\xi$ of the condensate (which we give here in units of the lattice spacing), 
we can restrict our analysis to the dynamics of
the $\pi$ mode and to the populations of modes with $q \approx 0$.
For the coherence and population of the $q=\pi$ mode we obtain
\begin{equation}\label{picoherence}
\partial_t \langle \hat{c}_\pi \rangle \simeq  - (i\delta_{\rm d}+ \Gamma +2 \kappa+4\kappa {\cal N}_0)  \langle \hat{c}_\pi \rangle -i \Omega_\pi,
\end{equation}
and
\begin{equation}
\partial_t \langle \hat{c}^\dag_\pi \hat{c}_\pi \rangle \simeq  -2\left[\Gamma+2\kappa +4 \kappa {\cal N}_0 \right] \langle \hat{c}^{\dag}_{\pi} \hat{c}_{\pi} \rangle + i \Omega_{\pi}^{\ast} \langle \hat{c}_{\pi} \rangle - i \Omega_{\pi} \langle \hat{c}_{\pi}^{\dag}\rangle. 
\end{equation}
Here we have defined 
\begin{equation}
{\cal N}_0 \equiv \frac{1}{L}\sum'_{q} n_q,
\end{equation}
with $\sum'_q$ 
excluding $q=\pi$, and used the Hartree-Fock decoupling
approximation, which for three operators reads $\langle
\hat{c}^{\dag}_{q_{1}+q_{2}-q} \hat{c}_{q_{1}} \hat{c}_{q_{2}} \rangle \simeq
n_{q_{1}} \delta(q_{2}-q) \langle \hat{c}_{q} \rangle + n_{q_{2}}
\delta(q_{1}-q) \langle \hat{c}_{q} \rangle$. For $\delta_{\rm d}=0$, the normalized
steady-state population ${\cal N}_{\pi} = \langle \hat{c}^\dag_\pi \hat{c}_\pi \rangle/ L$ of
the driven mode is then given by
\begin{equation}\label{pidensity}
{\cal N}_{\pi} =   \frac{|\Omega_\pi|^2/L} { \left[ \Gamma +2 \kappa(1+2{\cal N}_0)\right]^2 } .
\end{equation}
For the low-momentum modes we find
\begin{equation}
\partial_t n_q  \simeq   -\left[2\Gamma-  8\kappa{\cal N}_{\pi}+   \kappa(1+2{\cal N}_0+2{\cal N}_{\pi})q^2 \right] n_q+ 8 \kappa(1-q^2/4) {\cal N}_{\pi},
\end{equation}
and the steady-state solution can be approximately written in the form
\begin{equation} \label{HFsol}
n_{q} \simeq  \frac{n_0}{1 + \xi^{2} q^{2}},
\end{equation}
where
\begin{eqnarray}\label{n0xi}
n_0= \frac{8 \kappa {\cal N}_{\pi}}{2 \Gamma - 8 \kappa {\cal N}_{\pi}},\qquad \xi^{2} = \frac{\kappa(1+2{\cal N}_0+2{\cal N}_{\pi})}{2 \Gamma - 8 \kappa {\cal N}_{\pi}}+\frac{1}{4}.
\end{eqnarray}  

Equation (\ref{HFsol}) is a Lorentzian profile with broadening $1/\xi$.  This
scale plays the role of a correlation length since, in the
translationally-invariant case,
\begin{equation}
\langle \hat{c}^{\dag}_{\ell} \hat{c}_{\ell'} \rangle = \sum_{q,q'} e^{-i q \ell} e^{-i q' \ell'} \langle \hat{c}^{\dag}_{q} \hat{c}_{q'} \rangle \simeq \sum_{q} e^{-i q (\ell - \ell')} \langle \hat{c}^{\dag}_{q} \hat{c}_{q} \rangle \propto e^{-|\ell -\ell'| / \xi}.
\end{equation}
Here we have used that, in momentum space, the single-particle density matrix is
mainly occupied on the diagonal.  This approximation is in line with the
Hartree-Fock approximation used above and can be physically justified by the
fact that low-momentum modes are populated by photons which are scattered
incoherently from the $q=\pi$ mode.

\subsection{Discussion}

For the Lorentzian low-momentum distribution (\ref{HFsol}) we obtain ${\cal N}_0\simeq
n_0/\xi$, and we can interpret the quantity $L\times {\cal N}_0$ as the total number of
condensed photons, while the population $L \times {\cal N}_{\pi}$ serves as finite photon
reservoir.  In steady state, the values of ${\cal N}_{\pi}$, $n_0$ and $\xi$ can be
determined from the nonlinear Eqs. (\ref{pidensity}) and (\ref{n0xi}).  The
stability of the low-momentum modes requires that ${\cal N}_{\pi} <
\Gamma/4\kappa$, but for a driving strength above $\Omega_{\pi,{\rm crit}}$ the
value of ${\cal N}_{\pi}$ will approach this limit.  Therefore, in this regime
\begin{equation} \label{Nxi}
{\cal N}_0=\frac{n_0}{\xi}\simeq  \frac{(2\kappa+\Gamma)}{4\kappa} \left(\tilde \Omega_\pi-1\right), \qquad \xi \simeq \frac{\kappa {\cal N}_0 }{2\Gamma} \left( 1+ \frac{\Gamma}{2\kappa} +2 {\cal N}_0 \right),
\end{equation}
where $\tilde \Omega_\pi=|\Omega_\pi|/\Omega_{\pi,{\rm crit}}>1$. We see that
above the critical driving strength the number of photons in the low-momentum
modes increases linearly with $\tilde \Omega_\pi$. At the same time the
correlation length increases $\sim \tilde \Omega_\pi^2$, indicating an even
stronger tendency for photons to occupy the $q=0$ mode. Note that while in the
classical high-photon limit, $\Gamma\gg\kappa$, we obtain the ratio ${\cal N}_0/{\cal N}_{\pi} \simeq
(\tilde \Omega_\pi-1)$, in the opposite case, $\kappa\gg \Gamma$, we get an
additional enhancement factor, ${\cal N}_0/{\cal N}_{\pi} \simeq (2\kappa/\Gamma)(\tilde
\Omega_\pi-1)$. Therefore, in this regime, very pure, dilute photon
condensates with ${\cal N}_0\approx 1$ and $\xi\simeq (\kappa/4\Gamma) (\tilde
\Omega_\pi-1)\tilde \Omega_\pi\gg 1$ can be prepared. For typical
parameters, the correlation length ranges from a few to a few tens of cavities,
and can easily be tuned by adjusting the driving strength. This means that with
realistic settings of around ten cavities, the variation of the correlation
length could be studied in experiments.

Finally, we find it worthwhile to compare the steady-state distribution of the
non-equilibrium open system studied here to a thermal equilibrium gas of
massive bosons with finite chemical potential.  To do so, we compare the
distribution (\ref{HFsol}) with a Bose-Einstein distribution $n_q^{\rm
  BE}=1/(\exp\{(\varepsilon_q-\mu)/k_B T\} -1)$ (c.f. Fig.~\ref{figurenq}b), where $\varepsilon_q = -2J
\cos(q)$ in our case.  Considering small momenta, a low $q$ expansion of both
distributions gives the following effective temperature and
chemical potential for our photon condensate
\begin{eqnarray} \label{Tmu}
\frac{k_B T}{J} = \frac{1+n_0}{\xi^2}, \qquad
\frac{\mu}{J} = -2 - \frac{1+n_0}{\xi^2} \ln \left( 1+ \frac{1}{n_0} \right).
\end{eqnarray}

Note that the temperature $T$ defined in this way is different from $T_{\rm
  eff}$ associated to a finite $\kappa'$, and arises purely from the competition
between photon losses and equilibration processes in a driven system. Above
threshold, where $n_0,\xi \gg1$, we obtain the scaling $k_BT/J\simeq 4/\tilde
\Omega$ and $k_BT/J\simeq 2\Gamma/(\kappa\tilde \Omega) $ for the limits
$\Gamma\gg\kappa$ and $\Gamma\ll \kappa$, respectively. This confirms the
conclusion from above that, for $\kappa/\Gamma\rightarrow \infty$, arbitrarily
pure photonic condensates can be prepared with our scheme.

As a final remark, we point out that the detuning $\delta_{\rm d}$ of the driving
field, which enters as an energy offset in the cavity Hamiltonian
(\ref{Hamiltonianq}), affects the effective chemical potential $\mu$ only
indirectly via a modification of $\Omega_{\pi,{\rm crit}}$. In particular, and
in strong contrast to equilibrium systems, the sign of $\delta_{\rm d}$ does not play a
role in our non-equilibrium scenario and the chemical potential is related to
the strength, rather than to the detuning of the driving field, as seen from the combined Eqs.~(\ref{Tmu}) and (\ref{Nxi}). 
While for the present system this relation can be derived explicitly, it also suggests a
general way to think about the effect of driving fields on the stationary states
of open photonic quantum many-body systems.

\section{Conclusions and outlook} \label{ConclusionsSec}

In conclusion, we have analyzed a scheme to achieve condensation of microwave
photons in an open and driven array of superconducting cavities.  In particular,
we have shown how the coupling to superconducting qubits can be used to engineer
non-local and number-conserving dissipation processes for microwave photons,
which mimic the coupling to an effective low-temperature bath.  Under
state-of-the-art experimental conditions, these processes can exceed the
intrinsic losses in the system and produce stationary many-body states of
photons, which are independent of the details of the external driving fields. We
have proposed a basic experimental setup for demonstrating a stationary
condensate of photons in two or multiple coupled cavities and compared the
properties of such an out-of-equilibrium photon condensate to an equilibrium
scenario for massive bosons.

More generally, our work illustrates the potential of superconducting cavity
arrays for simulating various dissipative and out-of-equilibrium quantum many-body problems. 
The present example of photon condensation already shows how in
engineered and fully controlled systems, an appropriately designed coupling to a bath can lead to
non-trivial stationary states of an otherwise non-interacting photonic system,
and could provide new insights for related experiments in condensed-matter
systems.  Beyond this basic example, the proposed scheme for engineering
dissipation can be easily extended and combined with strong coherent
photon-photon interactions. Compared to original ideas described in the context
of cold atoms, superconducting circuits represent a complementary platform with, in many respects, additional flexibility 
to design non-standard dissipation processes, as well as to study open, non-equilibrium quantum many-body systems.

\section*{Acknowledgments}

The authors would like to thank P. Zoller for proposing the initial direction of this project and further insightful discussions. 
We also thank F. Marquardt for valuable feedback on this work. This work
was supported by the EU network AQUTE and the Austrian Science Fund (FWF)
through SFB FOQUS and the START grants Y 591-N16 (P.R.) and Y 581-N16 (S.D.).

\appendix

\section{Supplement to Section \ref{Results}}\label{AppendixB}

In this Appendix we extend Section \ref{Results}.  Specifically, we study
the case of incoherent pumping of the cavities and
analyze the stability of the steady-state solutions found in that section.

\begin{figure}
\begin{center}
\includegraphics[width=0.48\linewidth]{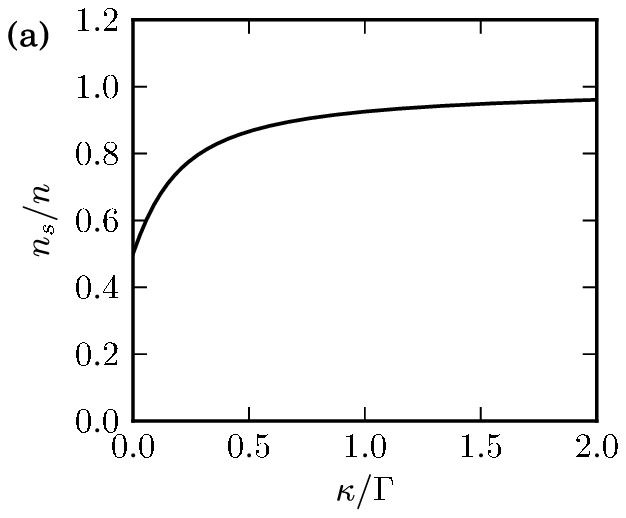}
\includegraphics[width=0.48\linewidth]{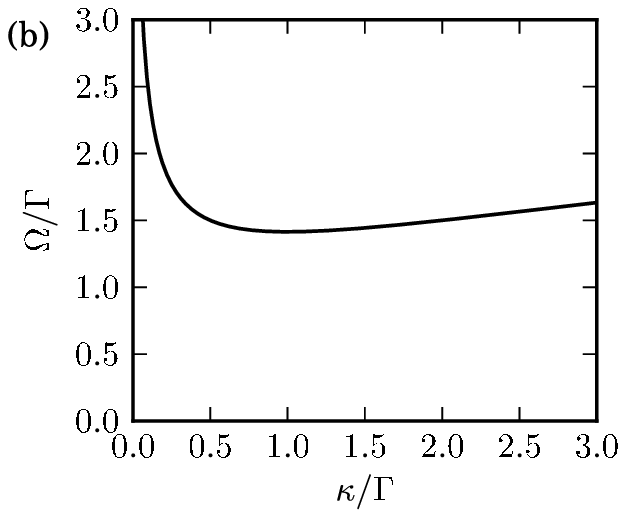}
\caption{\label{incoherent_pumping} a) Fraction of photons in the symmetric mode,
  $n_s/n\equiv\langle\hat{n}_s\rangle/n$ as a function of the dissipative rate
  $\kappa$, for incoherent pumping $p/\Gamma=0.5$. 
  b) Threshold given by Eq.~(\ref{threshold}), below which (\ref{solbelowthreshold}) is the only stable solution 
and above which (\ref{solabovethreshold}) is the only stable solution.}
\end{center}
\end{figure}

\emph{Incoherent pumping.}  It is important to notice that similar condensation
effects to those shown in Section \ref{Results} can be achieved under incoherent
pumping of the cavity modes.  Neglecting the term $\Omega \left(
  \hat{c}_a^{\dagger}+\hat{c}_a \right)$ in the model, and including
\begin{eqnarray}
{\cal L}_{\rm pump} = p {\cal D}[\hat{c}_s^{\dagger}] + p {\cal D}[\hat{c}_a^{\dagger}],
\end{eqnarray}
we arrive at the EOM
\begin{equation}
\begin{array}{l} \label{nEOM2cavities}
\partial_{t} \langle \hat{n}_{\rm s} \rangle = 2 p (\langle \hat{n}_{\rm s} \rangle + 1) - 2 \Gamma \langle \hat{n}_{\rm s} \rangle + 2 \kappa \langle \hat{n}_{\rm a} ( \hat{n}_{\rm s}  + 1) \rangle. \\
\partial_{t} \langle \hat{n}_{\rm a} \rangle = 2 p (\langle \hat{n}_{\rm a} \rangle + 1) - 2 \Gamma \langle \hat{n}_{\rm a} \rangle - 2 \kappa \langle \hat{n}_{\rm a} ( \hat{n}_{\rm s}  + 1) \rangle.
\end{array}
\end{equation}
Solving for the steady-state solution of these equations within the
semiclassical approximation $\langle\hat{n}_s\hat{n}_a\rangle \approx
\langle\hat{n}_s\rangle \langle \hat{n}_a \rangle$, we find the condition
$\frac{p}{\Gamma} = \frac{n}{n+2}$, being $n\equiv \langle \hat{n}_s
\rangle+\langle \hat{n}_a\rangle$ the total number of particles in the steady
state.  Substituting then $\langle
\hat{n}_a\rangle=n-\langle\hat{n}_s\rangle=\frac{2p}{\Gamma-p}-\langle\hat{n}_s\rangle$
in the first steady-state equation, for $p\neq\Gamma$, we find a smooth
crossover between an equal population of symmetric and antisymmetric modes for
$\kappa\ll p,\Gamma$, and condensation to the symmetric mode for $\kappa\gg
p,\Gamma$ (see Fig.~\ref{incoherent_pumping}a).  Interestingly, for $p\to\Gamma$,
however, all photons occupy the symmetric mode for all values of $\kappa\neq0$.
In the latter case, heating (incoherent pumping) compensates cooling, and only
the engineered dissipative mechanism survives.

\emph{Stability of the steady-state solutions.}  The stability of the different
solutions can be checked by linearization of the EOM around the steady-state
values (\ref{solbelowthreshold}) and (\ref{solabovethreshold}).  To this end we
linearize the EOM (\ref{meanfieldEOM2cavities}) as $s\to s_0 + \delta s$, $a\to
a_0 + \delta a$ (similarly for the conjugate fields), being $s_0$, $a_0$ the
steady-state solutions and $\delta s$, $\delta a$ representing fluctuations.
This gives $\frac{d}{dt} (\delta s, \delta a, \delta s^* \delta
a^*)_T = J (\delta s, \delta a, \delta s^* \delta a^*)_T$ with the $4\times
4$ Jacobian matrix $J$:
\begin{equation*}
\left ( \begin{array}{cccc}
-(\Gamma-\kappa|a_0|^2) & \kappa |s_0| a_0^* & 0 & \kappa |s_0| a_0 \\
-\kappa |s_0| a_0 & i\delta_c- [\Gamma+\kappa(|s_0|^2+1)] & -\kappa |s_0| a_0 & 0 \\
0 & \kappa |s_0| a_0^* & -(\Gamma-\kappa |a_0|^2) & \kappa |s_0| a_0 \\
-\kappa |s_0| a_0^* & 0 & - \kappa |s_0| a_0^* & -i\delta_c- [\Gamma+\kappa(|s_0|^2)]
\end{array} \right ),
\end{equation*}
whose eigenvalues will have negative real part in the region of stability.
Evaluating the eigenvalues of this matrix at the steady state solutions, we find
that the steady-state solutions presented in Section~\ref{Results} are stable in
the whole range of parameters.  In Fig.~\ref{incoherent_pumping}b we show 
the threshold $\Omega_{\rm crit}$ given by Eq.~(\ref{threshold}), below which (\ref{solbelowthreshold}) is the only stable solution 
and above which (\ref{solabovethreshold}) is the only stable solution.

\end{document}